# The architecture of the protein domain universe

Nikolay V. Dokholyan

Department of Biochemistry and Biophysics, The University of North Carolina at Chapel Hill, School of Medicine, Chapel Hill, NC 27599

**ABSTRACT**

Understanding the design of the universe of protein structures may provide insights into protein evolution. We study the architecture of the protein domain universe, which has been found to poses peculiar scale-free properties (Dokholyan et al., *Proc. Natl. Acad. Sci. USA* **99:** 14132-14136 (2002)). We examine the origin of these scale-free properties of the graph of protein domain structures (PDUG) and determine that that the PDUG is not modular, i.e. it does not consist of modules with uniform properties. Instead, we find the PDUG to be self-similar at all scales. We further characterize the PDUG architecture by studying the properties of the hub nodes that are responsible for the scale-free connectivity of the PDUG. We introduce a measure of the betweenness centrality of protein domains in the PDUG and find a power-law distribution of the betweenness centrality values. The scale-free distribution of hubs in the protein universe suggests that a set of specific statistical mechanics models, such as the self-organized criticality model, can potentially identify the principal driving forces of molecular evolution. We also find a gatekeeper protein domain, removal of which partitions the largest cluster into two large sub-clusters. We suggest that the loss of such gatekeeper protein domains in the course of evolution is responsible for the creation of new fold families.

## INTRODUCTION

The principles of molecular evolution remain elusive despite fundamental breakthroughs on the theoretical front [1-5] and a growing amount of genomic and proteomic data, over 23,000 solved protein structures [6] and protein functional annotations [7-9]. Like in other sciences, we can identify specific observables intrinsic to molecular evolution. In the context of protein sequence-structure-function relationships, graph-theoretical approaches proved to be effective to capture the nature of these relationships [4,10-14]. Thus, various network descriptors can be used as a choice of observables to study molecular evolution. Such descriptors are, for example, distributions of family members in protein families, connectivity of proteins in such networks and cliquishness. Theoretical efforts then can be directed to explain such descriptors.

Dokholyan et al. [10] proposed to study protein structural relations by constructing protein domain universe graphs (PDUG), in which nodes correspond to protein domains and edges connect pairs of nodes that correspond to structurally similar protein domains. The distribution of populations in protein families was explained from the theory of random graphs that requires no fitting parameters and does not rely on any evolutionary mechanisms. A peculiar imprint of evolution on protein structural space that, we believe, is non-trivial is the connectivity of protein domains in PDUG, the distribution $P(k)$ of connectivity $k$ is scale-free:

$$P(k) \propto k^{-\mu} \qquad (1)$$

where $\mu_{PDUG} \approx 1.6$. In contrast the distribution of connectivity in random graphs is Gaussian: $P_{RG}(k) \propto \exp(-(k-k_0)^2/2\sigma^2)$. Our quest to understand the mosaic structure of PDUG started from developing a simplified statistical mechanics model of protein domain evolution based on point mutations and gene duplication [10], which yielded the observed distribution Eq.(1) and exponent $\mu_0 \approx \mu_{PDUG} \approx 1.6$.

The success of a statistical mechanics model to explain the connectivity of PDUG proposed in Ref. [10] is undermined by its simplicity. Moreover, the model proposed in Ref. [10] does not take into account evolutionary pressure at the physical and biological levels, such as evolutionary selection of mutated proteins to be stable, to fold in a biologically reasonable time, and, possibly [15], to be biologically active. To test the rules of the model in Ref. [10], Deeds et al. [16] selected 3,500 conformations (the size of PDUG) out of $10^5$ completely enumerated compact structures of lattice 27-mer and showed that using selection rules similar to those proposed in Ref. [10], the distribution of connectivities $P(k)$ of 27-mers obeys Eq.(1) with the same exponent $\mu_{27mer} \approx \mu_{PDUG} \approx 1.6$.

From a physical point of view, England and Shakhnovich [17] suggested a link between the designability of a given fold, i.e. the number of sequences that are thermodynamically stable in the fold, and a structural characteristic of this fold – traces of even powers of a fold's contact matrix. Such a link unveiled a mechanism for thermophilic adaptation in bacterial genomes [18]: thermophilic proteins predominantly share higher designability folds than mesophilic ones. The important role of evolutionary pressure in preserving protein thermodynamic stability was emphasized by Dokholyan and Shakhnovich [15]. Tiana et al. [19] developed an evolutionary model of a lattice 36-mer in which stable protein structures are explicitly selected in Monte Carlo folding simulations. This selection procedure leads to a correlation between fold population and designability [17].

Here we dissect the PDUG architecture from the perspective of network organization. We ask whether the PDUG connectivity is the result of a specific distribution of clusters with homogeneous connectivity distribution within these clusters (*modularity*). Alternatively, the connectivity distribution may be identical, or *self-similar, to* the PDUG clusters and the entire PDUG. The scale-free organization of the PDUG has an important implication: it suggests a disparity in node connectivities over a broad range, which in turn implies that a few nodes are much more connected than others. These former nodes are hubs that are crucial to the small-world connectivity of the network. The question is then: what are those "central" hub nodes? To identify these hubs and characterize the network architecture, we employ a measure of the node centrality[20] in graph theory.

**IS THE PDUG MODULAR?**

The scale-free connectivity distribution in the PDUG may arise from two scenarios that imply different PDUG architectures. In the first scenario, the PDUG network is modular, i.e. the PDUG consists of clusters, in which the distribution of connectivities is Gaussian, but the distribution of such clusters is scale-free. In this scenario the connectivity distributions $P_c(k)$ in the PDUG clusters $c$ is Gaussian:

$$P_c(k) \propto \exp\left(-\frac{(k-\bar{k}_c)^2}{2\sigma_c^2}\right) \qquad (2)$$

where $\bar{k}_c$ and $\sigma_c$ are the average the standard deviation of the number of edges per node in the cluster $c$. For simplicity, let us also assume that the standard deviation is the same for all clusters, i.e. $\sigma_c \equiv \sigma = const$. Let $F_c(\bar{k}_c)$ denote the distribution of clusters in the PDUG with a given average number of nodes $\bar{k}_c$. The resulting distribution $P(k)$ is a convolution of these distributions:

$$P(k) = \sum_c P_c(k) F_c(\bar{k}_c). \qquad (3)$$

Then, Eq. (1) can be obtained if $F_c(\bar{k}_c)$ satisfies the following relation:

$$\int_0^\infty F_c(x) \exp\left(-\frac{(k-x)^2}{2\sigma^2}\right) dx \propto k^{-\mu}. \qquad (4)$$

The left hand side of Eq. (4) is a continuous form of the right hand side of Eq. (3). Eq. (4) is an integral equation, which in the limit of large $k$ has asymptotic solution

$$F_c(x) \propto x^{-\gamma}, \qquad (5)$$

where $\gamma = \mu$. Eq. (5) implies that the scale-free architecture of the PDUG results from the heterogeneous distribution of clusters, which, in turn, have a specific distribution with finite moments, such as averages and standard deviations.

In the second scenario, the connectivity distribution remains scale-free in PDUG clusters, i.e.

$$P_c(k) \propto k^{-\gamma}, \qquad (6)$$

in which case $\gamma = \mu$. We test whether the PDUG is modular by comparing the connectivity distributions of the PDUG sub-clusters and the PDUG. Thus, we ask whether the PDUG is modular or self-similar at all scales.

We compute the distributions $P_c(k)$ for three largest clusters in the PDUG and find that these distributions are not Gaussian (Fig. 1a). These distributions are broad: the connectivity distributions for the largest ($m_1$=123 nodes) and the third largest clusters reach maximal connectivity $k=26$ ($m_3$=53 nodes), while that for the second largest cluster reach maximal value of $k=61$ ($m_2=81$ nodes). The connectivity distribution of the second largest cluster is almost uniform for all values of $k$, while that for the first and the third largest cluster exhibit large dispersion in $k$. Fig. 1a suggests that the connectivity distributions posited in Eq. (2) are inconsistent with our observations, which means that the PDUG is, in fact, not modular. However, in order to provide a more quantitatively definite answer, we study the first and the second moments of the distributions Eq. (2) over various PDUG clusters.

Deciphering the distribution of values that follow power-law statistics is a challenging task due to the divergent nature of the power-law distribution: no matter what the value of the exponent is (e.g. $\mu$ in Eq. (1)), some moments of this distribution always diverge. The smaller the exponent $\mu$, the lower the moments diverge. Since natural systems are finite, such divergence comes as a finite size effect, i.e. the distribution is effectively modified by a truncation function of, for example, exponential origin[21]:

$$P(k) \propto k^{-\mu} f\left(\frac{k}{k_0}\right), \tag{7}$$

where $k_0$ is a cut-off value of the $P(k)$ distribution, which scales with the number of nodes in the system as some function of $N$. The leading term of the expansion of the $k_0(N)$ function is a power-law function $k_0 \propto N^\nu$.

In power-law distributions, the moments depend on the system size $N$. From Eq. (7) it follows, that the average and the standard deviation depend on $N$ as

$$\overline{k} \propto k_0^{-\mu+2} \propto N^{\nu(-\mu+2)}, \tag{8}$$

$$\sigma \propto k_0^{\frac{1}{2}(-\mu+3)}\left[1-\alpha k_0^{-\mu+1}\right]^{\frac{1}{2}} \propto N^{\frac{\nu}{2}(-\mu+2)}\left[1-\alpha N^{\nu(-\mu+1)}\right]^{\frac{1}{2}}, \tag{9}$$

where $\alpha$ is some constant. From Eqs. (8) and (9):

$$\sigma \propto \overline{k}^{\frac{1}{2}\frac{-\mu+3}{-\mu+2}}\left[1-\alpha \overline{k}^{\frac{-\mu+1}{-\mu+2}}\right]^{\frac{1}{2}}. \tag{10}$$

In contrast, the average and the standard deviation in a Gaussian distribution are not dependent on the system's size.

We find that both $\overline{k}$ and $\sigma$ scale (almost) linearly with $N$ and, accordingly, $\sigma$ scales linearly with $\overline{k}$ (Fig. 1b). Such behavior is consistent with Eqs. (7) – (10) only when exponents $\mu = \nu = 1$. In which case $\overline{k} \propto N$, $\sigma \propto N$, and $\sigma \propto \overline{k}$. This means that (i) the PDUG is not modular but self-similar at all scales, (ii) the observed exponent $\mu_{PDUG} \approx 1.6$ in Ref. [10] is an "effective" exponent due to the finite size effects that we cannot separate from the true power-law regime, and (iii) the true PDUG exponent is $\mu_{PDUG}^{true} = 1$.

This exponent is also consistent with that of the distribution of cluster populations $P(N)$, found[10] to scale as a power-law with the exponent 1.5. The distribution of the averages $P_a(\overline{k})$ is expected to scale the same as $P(N)$ since $P_a(\overline{k}) = P(N) \frac{dN}{d\overline{k}} \propto P(N)$. Although we find the power-law exponent of $P_a(\overline{k})$ to be 2 (Fig. 1c), the difference is believed to come from the finite size scaling of the $P_a(\overline{k})$ distribution.

The value $\mu_{PDUG}^{true} = 1$ of the power-law exponent has profound implications on the type of models that explain processes resulting in such distributions [22-25]. One such model is a model of self-organized criticality (SOC) proposed by Bak et al. [22]. In this model, the system that is out of equilibrium self-organizes (no tunable model parameters are required) into a dynamic unstable attractor. Such a model may be applicable to evolution of proteins: a family of proteins may start via gene duplication and point mutations as a structural derivative *A* of a protein *B* representing some other family. Next, protein *A* generates a family via gene duplication and point mutations [15] while diversification is favored by the selection process. Protein *A* serves as a hub for a new family. The fluctuations in selection processes may result in diverse connectivities between proteins.

The scale-free connectivity distribution may also have a structural origin: some mutations have little effect on protein structure, while others drastically alter the three-dimensional protein conformation. Unfortunately, large-scale predictions of the effects of mutations on protein structures are not currently feasible. Thus, it is difficult to probe this scenario *in silico*.

**BETWEENNESS CENTRALITY**

The hubs play an important role in the network architecture. These special nodes are central to a network as a large number of geodesics, or minimal path, pass through these nodes. To characterize these hubs we employ a measure of the node betweenness centrality proposed by Freeman [20] (Methods).

The PDUG was built in Ref.[10] by first constructing a weighted graph consisting of nodes representing protein domains, edges connecting those nodes that are structurally similar, and weights that are a measure of protein domain structural similarity. The measure of structural similarity are DALI[26] Z-scores. The original weighted graph is then partitioned by removing all edges that have Z-scores smaller than a specific $Z_{min}$. After partitioning, the weights are discarded and the final graph is denoted as PDUG. The value of $Z_{min}$ is unambiguously chosen so that it corresponds to the midpoint of a transition of the dependence of the number of nodes in the largest cluster as a function of $Z$[10]. It was found that $Z_{min}=9$. The largest cluster of PDUG mostly represents the Rossman fold and contains 123 protein domains (Fig. 2b), the second largest cluster represents the TIM-barrel fold and contains 81 protein domains (Fig. 2c), and the third largest cluster represents the Immunoglobulin-like beta-sandwich and contains 53 protein domains (Fig. 2d).

We determine the betweenness of each protein domain in the three largest clusters (e.g. Fig. 3a). We find that for these three clusters the distributions of the betweenness are highly heterogeneous and are well-fit by a power law function:

$$P_B(B) \propto B^{-\eta}, \tag{11}$$

where $\eta \approx 1$. Strikingly, the exponent $\eta$ is the same as $\mu_{PDUG}^{true}$. In contrast, the betweenness distribution $P_B^{RG}(B)$ of random graphs (Methods) exhibits distinctly different behavior (Fig. 3b):

$$P_B^{RG}(B) \propto \exp(-B/\bar{B}), \tag{12}$$

where $\bar{B} \approx 2\times 10^4$ is the average value of betweenness in random graphs. In the case of random graphs, we generally obtain larger clusters than in the PDUG, which explains the larger values of *B* in random graphs than in the PDUG (Fig. 3).

The scale-free distribution of betweenness in the PDUG has very non-trivial implications on our understanding of protein evolution. In one scenario, in the course of evolution a central node "gives birth" (through gene duplication and point mutations) to a world of protein domains. Evolution proceeds by such subsequent formation of these central nodes and their corresponding worlds. These hubs exist on all scales, consistent with the self-similar PDUG architecture. The scale-free "hub" architecture of the PDUG may be responsible for the connectivity properties of the protein universe. However, due to the lack of a simple relationship between the betweenness of a node and its connectivity, it is challenging to mathematically relate corresponding distributions $P(k)$ and $P_B(B)$.

Interestingly, in the case of the largest cluster (Fig. 2a), three protein domains 1cex_1 (cutinase from *Fusarium solani*), 1qpzA_5 (N-terminal domain purine repressor (PurR) from *Escherichia coli*), and 1gca_2 (galactose/glucose-binding protein from *Salmonella typhimurium*) have the largest betweenness among all domains belonging to the largest cluster: approximately 28% of the shortest paths pass through these three nodes. The domain 1cex_1 is a "gatekeeper" to a large sub-cluster of the largest cluster: removal of this node will result in separation of the two largest sub-clusters (Fig. 2b). It is possible that the disappearance of such gatekeeper protein domain results in the formation of new protein families. Therefore, uncovering the evolutionary origin of this domain may offer insights into formation of protein domain worlds.

Cutinases (e.g. 1cex_1) are hydrolytic enzymes that degrade cutin, a polyester composed of hydroxy and epoxy fatty acids, that serves as a protective shield of aerial plants against pathogens entry[27]. Cutin degradation is the first step of plant infection and it is exploited by fungi that express cutinase to invade plants. This unique mechanism of plant protection and infection may trace back to the very origin of plants. Co-evolution of plants and pathogens may have spurred an observed spread of domains in the largest cluster of the PDUG.

**CONCLUSION**

Insights into the evolution of proteins and, further, of organisms, may likely come from an understanding of the architecture of protein structural space [4,28-32] rather than that of sequence space [10,15]. We, therefore, scrutinize the architecture of the PDUG that was found to have a scale-free connectivity distribution. We find that the PDUG is not modular, i.e. it does not consist of modules with uniform properties, but rather is self-similar at all scales. We further find that the true power-law exponent $\mu_{PDUG}^{true} = 1$ rather than $\mu_{PDUG} \approx 1.6$, as was previously determined[10]. The discrepancy between the true and effective exponents is postulated to be due to the finite size of the PDUG.

We further characterize the PDUG architecture by studying the properties of the hub nodes that are responsible for the scale-free connectivity of the PDUG. The power-law distribution of the betweenness centrality of protein domains manifests the scale-free organization of PDUG. The scale-free architecture of the PDUG, especially the power-law exponent $\mu_{PDUG}^{true} = 1$, suggests that a set of specific statistical mechanics models, such as SOC, may explain such distributions. These models may provide insights into the most fundamental processes that exist in Nature and that guide the evolutionary course of proteins and organisms.

We also find a gatekeeper protein domain 1cex_1, removal of which partitions the largest cluster into two large sub-clusters. We suggest that these gatekeeper protein domains, or the loss of them in the course of evolution, is responsible for the creation of new fold families. We argue that the origin of the gatekeeper domains may arise from co-evolution of organisms.

**METHODS**
**Betweenness centrality measure**

Let $G = \{V, E\}$ be an undirected unweighted graph, where $V$ is a set of nodes and $E$ is a set of edges. A *path* between nodes $s$ and $t$ is defined as a set of edges that connect nodes $s \in V$ and $t \in V$ on $G$, the length of which is determined as the number of edges on the path. The

*minimal path* is defined as the path of minimal length. Let $\sigma_{st}$ be the number of minimal paths between nodes $s \in V$ and $t \in V$, and $\sigma_{ss} \equiv 1$. Let $\sigma_{st}(v)$ denote the number of minimal paths between nodes $s \in V$ and $t \in V$ that pass through the node $v$. The *betweenness centrality* $B(v)$ is defined as

$$B(v) = \sum_{s \neq t \neq v \in V} \frac{\sigma_{st}(v)}{\sigma_{st}}. \qquad (13)$$

High betweenness of a node indicates that a large number of minimal paths pass through this node. The betweenness of a node $v$ is therefore a measure of participation of this node in connectivity of the graph. We implement a fast algorithm for betweenness computation proposed by Brandes [33]. In order to maintain the information on the actual values of $B$ we do not normalize $B$ to range between 0 and 1 as is often done.

**Construction of random graphs**

Random graphs are constructed by reshuffling the PDUG edges between randomly selected pairs of nodes. We perform the reshuffling operation $10^6$ and $10^7$ times and find that the distributions $P_B^{RG}(B)$ in both cases are almost identical, suggesting that reshuffling has effectively eliminated the PDUG architecture.

**ACKNOWLEDGEMENT**

We would like to thank Shantanu Sarma for implementing the Brandes [33] algorithm for fast calculation of the betweenness centrality. This work is supported by the UNC/IBM Junior Faculty Award.

# FIGURE CAPTIONS

**Figure 1.** (a) The connectivity distributions $P_c(k)$ of the three largest PDUG clusters consisting of 123, 81, and 53 members correspondingly. All of these three distributions have large first and second moments, which suggest that the underlying distributions are not Gaussian. (b) Scatter plots of three sets of data $\bar{k}$ versus $N$ (circle), $\sigma$ versus $N$ (square), and $\sigma$ versus $\bar{k}$ (diamond). These three scatter plots are well fit by linear functions. The slopes, linear regression correlation coefficients, and their corresponding $p$-values for a power-law fit on a double logarithmic scale are correspondingly (slope ≈ 0.9, $R$ ≈ 0.87, $p$ ≈ $10^{-31}$), (slope ≈ 1.0, $R$ ≈ 0.99, $p$ ≈ $10^{-21}$), and (slope ≈ 1.0, $R$ ≈ 0.98, $p$ ≈ $10^{-21}$). The numbers of points in these fits are 179, 93, and 93, correspondingly. The difference in number of points is due to exclusion of points that feature $\sigma = 0$ in $\sigma$ versus $N$, and $\sigma$ versus $\bar{k}$ fits. The straight line with slope $= 1$ is drawn as a visual guide. The regression slopes ≈1 imply almost linear relations between fitted values: $k \propto N^{0.9}$, $\sigma \propto N$, and $\sigma \propto \bar{k}$. (c) The probability distribution of the average values of node connectivities in the PDUG clusters. The slope, linear regression coefficient, and the $p$-value are slope ≈ $-2$, $R$ ≈ $-0.9$, and $p$ ≈ 0.01. The regression slopes imply the following relation between fitted values: $P(\bar{k}) \propto \bar{k}^{-2}$.

**Figure 2.** (a) The values of betweenness centrality of each individual node in the largest cluster of the PDUG. Three protein domains 1cex_1, 1qpzA_5, and 1gca_2, marked as red triangles, are most central to the largest PDUG cluster. (b), (c), and (d) The graph of representations of the first, the second and the third largest clusters correspondingly. A representative structure accompanies each of these graphs. In (b) we show the structures of the three protein domains 1cex_1, 1qpzA_5, and 1gca_2.

**Figure 3.** (a) The probability distributions of the betweenness centrality values of the PDUG nodes in the first (circle), the second (square), and the third (diamond) largest clusters on a double-logarithmic plot. All these distributions are similar and inversely proportional to $P_B(B) \propto 1/B$. The slopes, linear regression coefficients, and the $p$-values of the linear regression are correspondingly (slope $= -1.1$, $R = -0.96$, $p = 10^{-4}$), (slope $= -1.2$, $R = -0.92$, $p = 10^{-3}$), and (slope $= -1.0$, $R = -0.95$, $p = 10^{-3}$). The straight line with the slope $= -1$ is drawn as a visual guide. The regression slopes imply the following relations between fitted values: $P_B(B) \propto B^{-1.1}$, $P_B(B) \propto B^{-1.2}$, and $P_B(B) \propto B^{-1.0}$. (b) The probability distributions of the betweenness centrality values of the random graph nodes in the largest cluster on semi-logarithmic plot. The slope, linear regression coefficient, and the $p$-value of the linear regression are correspondingly slope $= -2 \times 10^4$, $R = -0.97$, and $p = 10^{-5}$. The regression slopes imply the following relation between fitted values: $P_B^{RG}(B) \propto \exp(-B/2 \times 10^4)$.

# FIGURES

# Figure 1

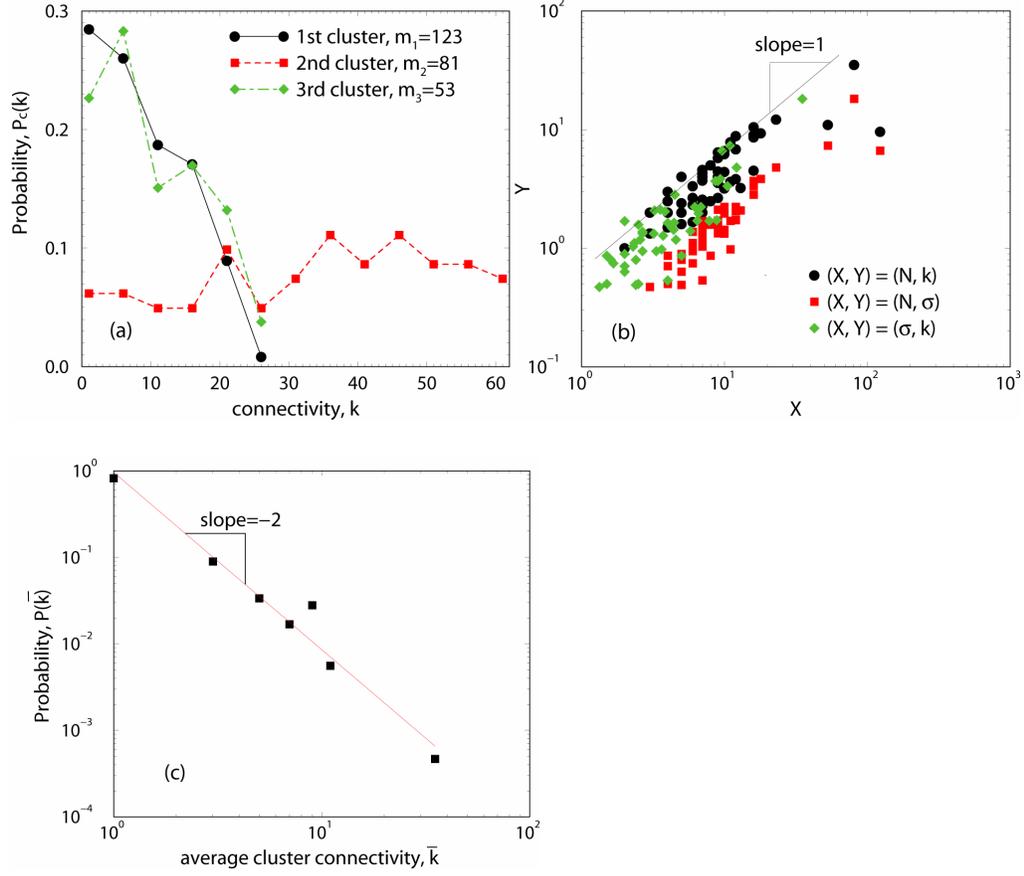

**Figure 2**

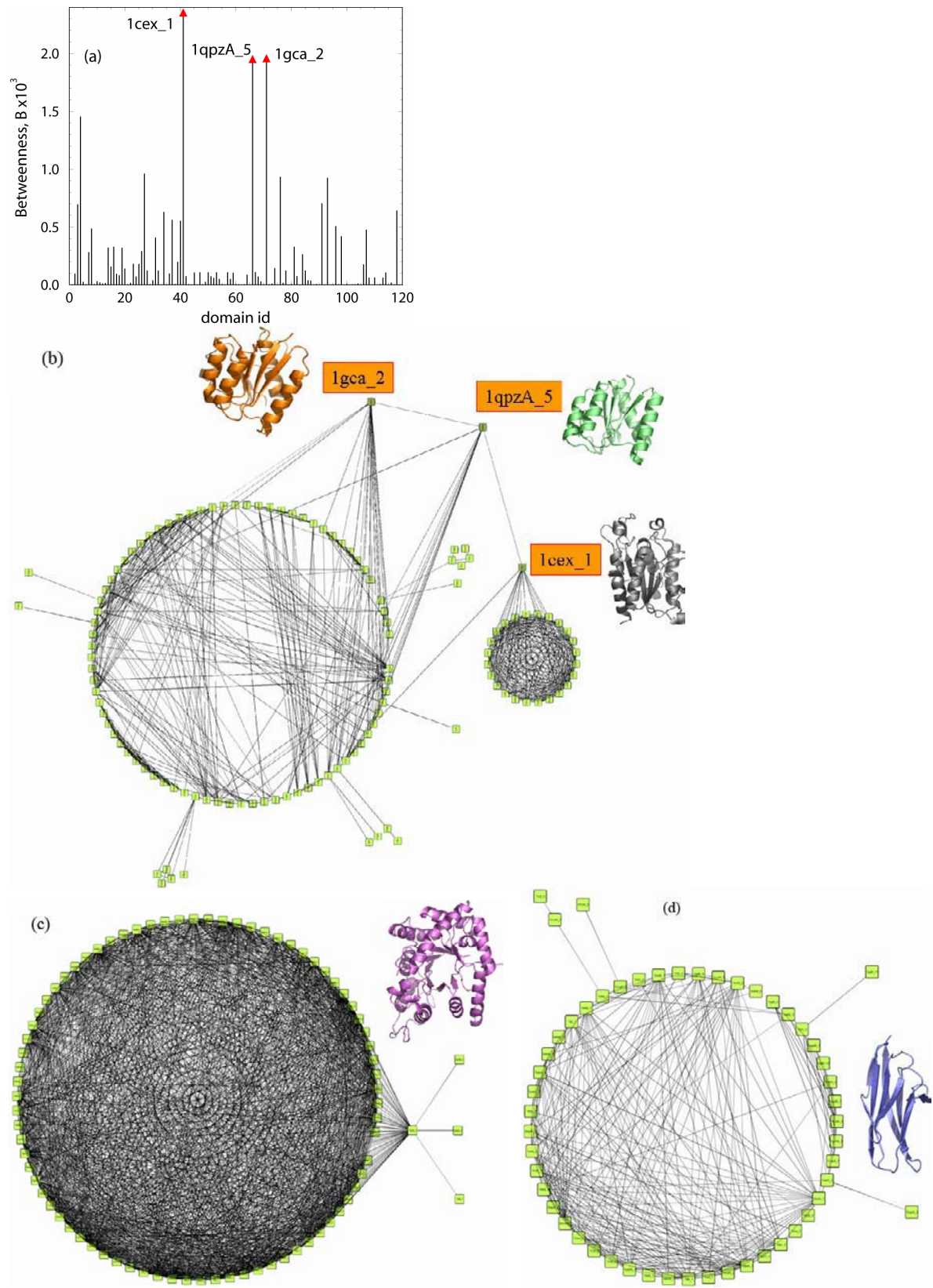

**Figure 3**

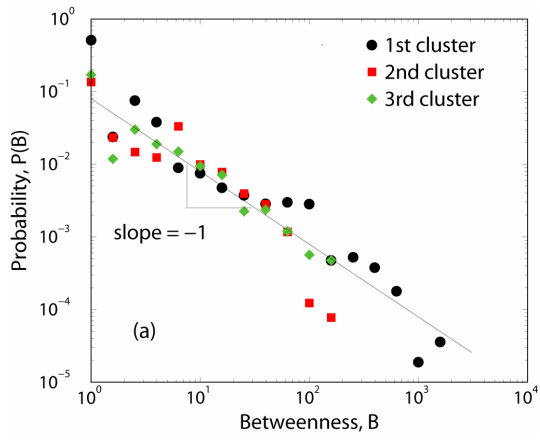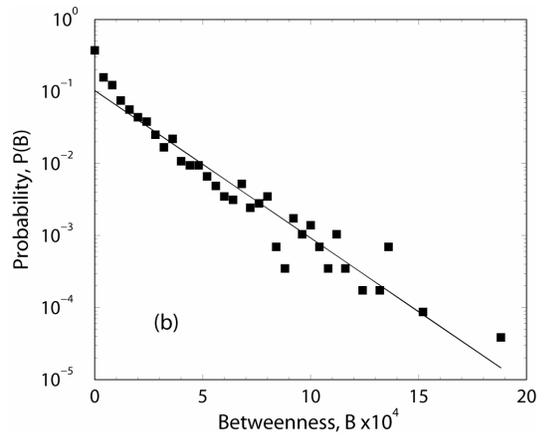